\documentclass[superscriptaddress,prl,preprintnumbers,twocolumn,nofootinbib]{revtex4-1}

\usepackage[utf8]{inputenc}
\usepackage{amsmath,amssymb,amsfonts}
\usepackage{bm, slashed, soul}
\usepackage{graphicx}
\usepackage{hyperref}
\usepackage{multirow}
\usepackage{comment}
\usepackage{color}
\definecolor{red}{rgb}{0.9, 0,0}

\newcommand{\lmk}{\left(}  
\newcommand{\rmk}{\right)}

\newcommand{\del}{\partial}

\newcommand{\beq}{\begin{eqnarray}}
\newcommand{\eeq}{\end{eqnarray}}

\newcommand{\be}{\begin{equation}}
\newcommand{\ee}{\end{equation}}

\newcommand{\bea}{\begin{array}}
\newcommand{\eea}{\end{array}}

\newcommand{\abs}[1]{\left\vert #1 \right\vert}
\newcommand{\eq}[1]{Eq.~(\ref{#1})}

\def\beq#1\eeq{\begin{align}#1\end{align}}

\begin{document}

\title{Charged black holes in non-linear Q-clouds with O(3) symmetry}

\author{Jeong-Pyong Hong}
\affiliation{Center for Theoretical Physics, Department of Physics and Astronomy, Seoul National University, Seoul 08826, Korea}
\author{Motoo Suzuki}
\affiliation{Institute for Cosmic Ray Research, The University of Tokyo, 5-1-5 Kashiwanoha, Kashiwa, Chiba 277-8582, Japan}
\author{Masaki Yamada}
\affiliation{Institute of Cosmology, Department of Physics and Astronomy, 
Tufts University, 574 Boston Avenue, Medford, MA 02155, U.S.A.}

\date{\today}

\begin{abstract}
We construct charged soliton solutions around spherical charged black holes with no angular momentum in asymptotically flat spacetime. These solutions are non-linear generalizations of charged scalar clouds, dubbed Q-clouds, and they do not contradict the non-existence theorem for free (linear) scalar clouds around charged black holes. These solutions are the first examples of O(3) solutions for Q-clouds around a non-extremal and non-rotating BH in the Abelian gauge theory. We show that a solution exists with an infinitely short cloud in the limit of extremal black holes. We discuss the evolution of Q-cloud in a system with fixed total charge and describe how the existence of Q-clouds is related to the weak-gravity conjecture. The reason that the no-hair theorem by Mayo and Bekenstein cannot be applied to the massive scalar field is also discussed.

\end{abstract}

\maketitle
\preprint{}


{\bf Introduction.--}
The strong gravitational effects of black holes (BHs) 
allow us to study the connections between theories of gravitation and quantum-field theory. 
One of the most important implications of the conjunction of quantum field theory and BHs is Hawking radiation~\cite{Hawking:1974rv}, which is emitted because of the creation of particle pairs near the surface of BHs. 
Several conjectures 
have been proposed through the analysis of BHs~\cite{tHooft:1993dmi, Susskind:1994vu, Bousso:1999xy, Ryu:2006bv}, and these serve as guides toward important insights into fundamental theories of physics. 
In particular, 
the weak-gravity conjecture~\cite{ArkaniHamed:2006dz} addresses the inconsistency of theory with BH remnants~\cite{Giddings:1992hh, tHooft:1993dmi, Susskind:1995da, Bousso:2002ju} 
and the non-existence of global symmetries in string theory~\cite{Banks:2010zn}. 
The latter fact is consistent with the no-hair theorem, which states that 
a BH can be described with only a finite number of parameters, like its mass, angular momentum, and gauge charge.

In relation to quantum field theory, 
the possibility of the existence of scalar cloud around a BH is an interesting avenue for research. 
As scalar fields can construct solitonic objects through self-interactions or gravitational interactions, 
BHs may have an extended scalar cloud outside their event horizons. 
Much effort has been devoted to finding such a stable solution around a BH, 
and there exist many solutions around rotating BHs~\cite{Hod:2012px, Hod:2012zza, Herdeiro:2014goa, Hod:2014baa, Benone:2014ssa, Huang:2016qnk, Huang:2017whw}. This is because the angular momentum prevents the field from being absorbed into the BH. One may expect that a Coulomb repulsion can play the same role for a charged BH. 
However, free-field theories include a non-existence theorem for scalar clouds around a non-rotating charged 
BH~\cite{Furuhashi:2004jk, Hod:2013eea, Hod:2013nn, Hod:2015hza} 
(see also Refs.~\cite{Hod:2012zza, Hod:2016yxg, Hod:2016iri, Huang:2016qnk}). This can be understood by noting that 
both gravitational and electric potentials behave as $\sim 1/r$ (at least at a large distance from a BH) 
while the effective potential due to the angular momentum behaves as $\sim1/r^2$. 
We cannot make a local minimum by using the former two potentials 
while we can make the one by adding the latter potential.

In this paper, we demonstrate the first examples of O(3) solutions for Q-clouds around a non-extremal and non-rotating BH in the Abelian gauge theory with a complex scalar field, which are realized by introducing the self-interaction of the scalar field.%
\footnote{
Yang-Mills hair around a charged BH was studied in Refs.~\cite{Bartnik:1988am, Volkov:1989fi, Kuenzle:1990is, Bizon:1990sr, Herdeiro:2017oxy, Maki:2019kgw}, Proca clouds were studied in Ref.~\cite{Sampaio:2014swa}, 
and scalar hair around an extremal charged BH was studied in Refs.~\cite{Murata:2013daa, Angelopoulos:2018yvt}. 
For Q-clouds around a Kerr (rotating) BH, see Refs.~\cite{Herdeiro:2014pka, Herdeiro:2015tia, Herdeiro:2017oyt}.
}
We consider a charged BH and introduce an attractive self-interaction in the charged scalar field. 
The attractive self-interaction of the scalar field allows the flat spacetime 
to form a localized condensate, known as Q-ball~\cite{Coleman:1985ki, Lee:1988ag, Kusenko:1997si}. 
This solution may hold even in the presence of a BH at the center of the Q-ball, 
which state is dubbed a Q-cloud around a BH. 
The Q-cloud may be unstable in this case as the BH absorbs the scalar field at its horizon. 
However, 
the gauge interaction prevents the charged scalar field from being
absorbed into the charged BH if the Q-cloud and the BH have charges of the same sign. 
We find that 
a stable solution can be constructed when these effects are in balance. 
We also discuss the evolution of the Q-cloud and 
show that 
an initially near-extremal BH evolves into a non-extremal BH with Q-cloud. 
The existence of such Q-cloud is supported by the 
weak-gravity conjecture. 

\vspace{0.1cm}
{\bf Charged BH and O(3) Q-cloud.--}
We consider a Reissner-Nordstr\"{o}m BH, which is a non-rotating charged BH. It is 
described by the following metric:
\begin{align}
ds^2=-\frac{\Delta}{r^2}dt^2+\frac{r^2}{\Delta}dr^2+r^2d\theta^2+r^2\sin^2\theta d\phi^2.
\end{align}
We use the Planck unit $G=c=\hbar=1$ throughout this paper. 
We define
\begin{align}
\Delta\equiv r^2-2M_{\rm BH}r+Q_{\rm BH}^2,
\end{align}
where $M_{\rm BH}$ and $Q_{\rm BH}$ are the mass and charge of the BH, respectively.
The horizons become the zeroes of $\Delta$, which are given as 
\begin{align}
r_\pm\equiv M_{\rm BH}\pm\sqrt{M_{\rm BH}^2-Q_{\rm BH}^2}.
\end{align}
The charge of the BH induces an electrostatic potential outside the horizon.

We introduce a complex scalar field $\Phi$ that has charge $q$ under the same Abelian gauge symmetry:
\begin{align}
\mathcal{L}=(\nabla^\mu+iqA^\mu)\Phi^\ast(\nabla_\mu-iqA_\mu)\Phi-V(\Phi),
\end{align}
where $\nabla^\mu$ denotes the covariant derivative for Reissner-Nordstr\"{o}m metric, $A_\mu$ is U(1) gauge field, and $V(\Phi)$ is a potential of the scalar field specified later. 
This scalar field induces an electrostatic potential $A_0$ in the outer region. 
We denote the energy and number of the scalar field as $E_\phi$ and $N_\phi$, respectively, where 
\begin{align}
N_\phi=-2 \int d^3x\frac{r^2}\Delta\text{Im}\left[\Phi^\ast(\partial_0-iqA_0)\Phi\right].
\end{align} 
The electric charge of the scalar field is given by $Q_\phi = q N_\phi$. 
After adopting the following ansatz,
\begin{align}
\Phi(x)=\frac{1}{\sqrt{2}}\phi(r) e^{-i \omega_\phi t},
\end{align}
which is motivated by the Q-ball solution in the flat spacetime,
we obtain the following equations for $\phi$ and the zeroth component of gauge field $A_0$: 
\begin{align}
&\Delta\frac{d}{dr}\left(\Delta\frac{d\phi}{dr}\right)+r^4 g^2\phi-\Delta r^2V'(\phi)=0,
\label{eq}\\
&r^2\frac{d}{dr}\left(r^2\frac{dg}{dr}\right)-\frac{r^6}{\Delta} q^2g\phi^2=0,
\label{eqfull}
\end{align}
where we define $g\equiv\omega_\phi + qA_0$.
The boundary conditions are given by 
\begin{align}
&\phi' (r_+) = V'(\phi(r_+))\frac{r_+^2}{r_+-r_-},\qquad 
\phi(\infty) = 0,\\
&g(r_+)=0,\qquad 
g(\infty)= \omega_\phi.
\end{align}

In this letter, we consider the case in which $E_\phi \ll M_{\rm BH}$ and $Q_\phi \ll Q_{\rm BH}$ 
so that we can treat the metric and the U(1) gauge field as the background. 
Then, the gauge-field background is given by $A_0 = - Q_{\rm BH} / r$ 
and 
the equations of motion reduce to \eq{eq} with $g = \omega_\phi - q Q_{\rm BH} / r$.

We are only interested in stationary solutions, so $\omega_\phi$ must be equal to 
\begin{align}
\omega_c\equiv \frac{qQ_{\rm BH}}{r_+}. 
\label{omega_c}
\end{align}
Otherwise, the above equation asymptotically approaches 
\begin{align}
\frac{d^2\phi}{dr_\ast^2}+\left(\omega_\phi-\omega_c\right)^2\phi\sim0,
\end{align}
near the horizon, with $dr_\ast/dr\equiv r^2/\Delta$, which gives an incoming or outgoing (i.e., not stationary) wave solution, 
$\phi(r)\sim e^{-i(\omega_\phi-\omega_c)r_\ast}$,
along with the factor $e^{-i\omega_\phi t}$. 
The stationary condition, $\omega_\phi = \omega_c$, is known to be at the threshold for superradiance~\cite{Bekenstein:1973mi, Herdeiro:2014goa}.

For Q-balls in flat spacetime, the phase velocity $\omega_\phi$ is equal to $d E_\phi / d N_\phi$ and can be identified as the chemical potential of the Q-ball~\cite{Gulamov:2013cra}. 
Therefore if $\omega_\phi$ is smaller than the mass of $\Phi$ in vacuum, 
the energetically favored behavior is for a particle to be localized to form a Q-ball. 
In the presence of a BH at the center of a Q-cloud, a U(1) gauge interaction and a charged BH are needed to construct a stationary solution. 
The U(1) gauge interaction prevents the scalar field from being absorbed into the BH 
if the BH and Q-cloud have charges of the same sign. 
This behavior can also be understood from \eq{omega_c}: $\omega_c$ vanishes if the BH has no charge 
and cannot be equal to $\omega_\phi$.

\vspace{0.1cm}
{\bf Examples of Q-cloud.--}
We shall next discuss the properties of Q-cloud, 
specifying the scalar potential. 
We consider the case in which $V(|\Phi|)$ is given by a polynomial potential: 
\begin{align}
 V(\Phi)=\mu^2|\Phi|^2 - \lambda |\Phi|^4 + A |\Phi|^6. 
\end{align}
We assume $A > \lambda^2/4\mu^2$ so that $\Phi = 0$ is a true vacuum.

Let us begin to consider the limiting case in which 
the gauge charge is vanishingly small and the Q-ball radius is considerably larger than the BH radius. 
The gravitational effect of the BH (i.e., the change of the metric in the presence of BH) is negligible though the regularity condition Eq.~(\ref{omega_c}) on the phase velocity must be satisfied, no matter how large the Q-ball radius is. 
In this case, we can construct Q-balls just as we do in flat spacetime 
with a condition of $\omega_\phi = \omega_c$. 
In the thin-wall limit of the Q-ball~\cite{Coleman:1985ki}, 
the phase velocity $\omega_\phi$ 
and the Q-ball radius $R_Q$ (i.e., the radius of the thin wall) are given by 
\beq
 \omega_\phi \simeq 
 \mu \sqrt{1 - \frac{\lambda^2}{4A \mu^2} }, 
 \qquad 
 R_Q \simeq \lmk \frac{3 A Q_\phi}{4 \pi q \lambda \omega_\phi} \rmk^{1/3}. 
 \label{thin_wall_Q-ball}
\eeq
The amplitude of the scalar field at the center of the Q-ball is determined by 
minimizing $V(\phi) / \phi^2$ and is almost independent of $Q_\phi$ in the thin-wall limit~\cite{Kusenko:1997ad}. 
For a smaller and thicker Q-ball, $\omega_\phi$ is larger than this value but is smaller than $\mu$. 
So there exists a Q-ball solution 
only when $\mu \sqrt{1 - \frac{\lambda^2}{4 A \mu^2} } < \omega_\phi < \mu$. 
In the presence of a BH, 
$\omega_\phi$ must be equal to $\omega_c$ 
so that no energy flows at the BH surface. 
Thus, there exists a Q-cloud solution when $\mu \sqrt{1 - \frac{\lambda^2}{4 A \mu^2} } < \omega_c < \mu$. 
In other words, 
we can always construct an O(3) Q-cloud around a charged BH 
if there exists a large Q-ball solution in flat spacetime 
and if this solution satisfies $\omega_\phi = \omega_c$.

When the Q-cloud size, $R_Q$, is comparable to the BH size, $r_+$, 
the effect of the Reissner-Nordstr\"{o}m metric is important and 
the equation (\ref{eq}) can only be solved numerically. The shooting method can be used for this solution. 
The field value at the surface of the BH, $\phi_0$, is chosen in such a way that $\phi(r)$ approaches $0$ for the range $r \to \infty$. 
The unknown parameters we should specify are $M_{\rm BH}$, $Q_{\rm BH}$, $\mu$, $q$, 
and parameters that govern the self-interaction. 
Here we note that 
equation \eq{eq} does not change with the following rescaling: 
\beq
 M_{\rm BH} \to c M_{\rm BH} , \quad Q_{\rm BH} \to c Q_{\rm BH}, \quad r \to c r, 
 \nonumber
 \\
 \omega_\phi \to \omega_\phi /c, \quad \mu \to \mu / c, \quad q \to q /c, 
 \quad 
 \phi \to \phi /c. 
\eeq
We use invariant combinations under this rescaling, such as $Q_{\rm BH} / M_{\rm BH}$, $r / M_{\rm BH}$, 
$\omega / \mu$, $q / \mu$, and $\phi / \mu$, 
to show the numerical results. 
Note that $E_\phi$ and $Q_\phi$ can always be made much smaller than $M_{\rm BH}$ and $Q_{\rm BH}$, respectively, 
by choosing a small value of $c$ in the rescaling. Therefore, there always exists a parameter space in which 
back-reactions of the Q-ball to the metric and gauge potential are negligible. 
We can also rescale $\phi$ so that $\lambda = 1$ without losing generality. 
We take $A = \lambda^2 / 3 \mu^2$ as an example, which is chosen as the largest value under the condition that the potential has a local minimum only at the origin. 
We numerically determine the size of the Q-ball by identifying the radius at which $90\%$ of the Q-ball charge is enclosed. We adopt this definition for numerical calculations since it can be used for general numerical profile, while it roughly coincides with Eq.~(\ref{thin_wall_Q-ball}) for thin-wall profile.

\begin{figure}
\centering
  \includegraphics[width=0.9\linewidth]{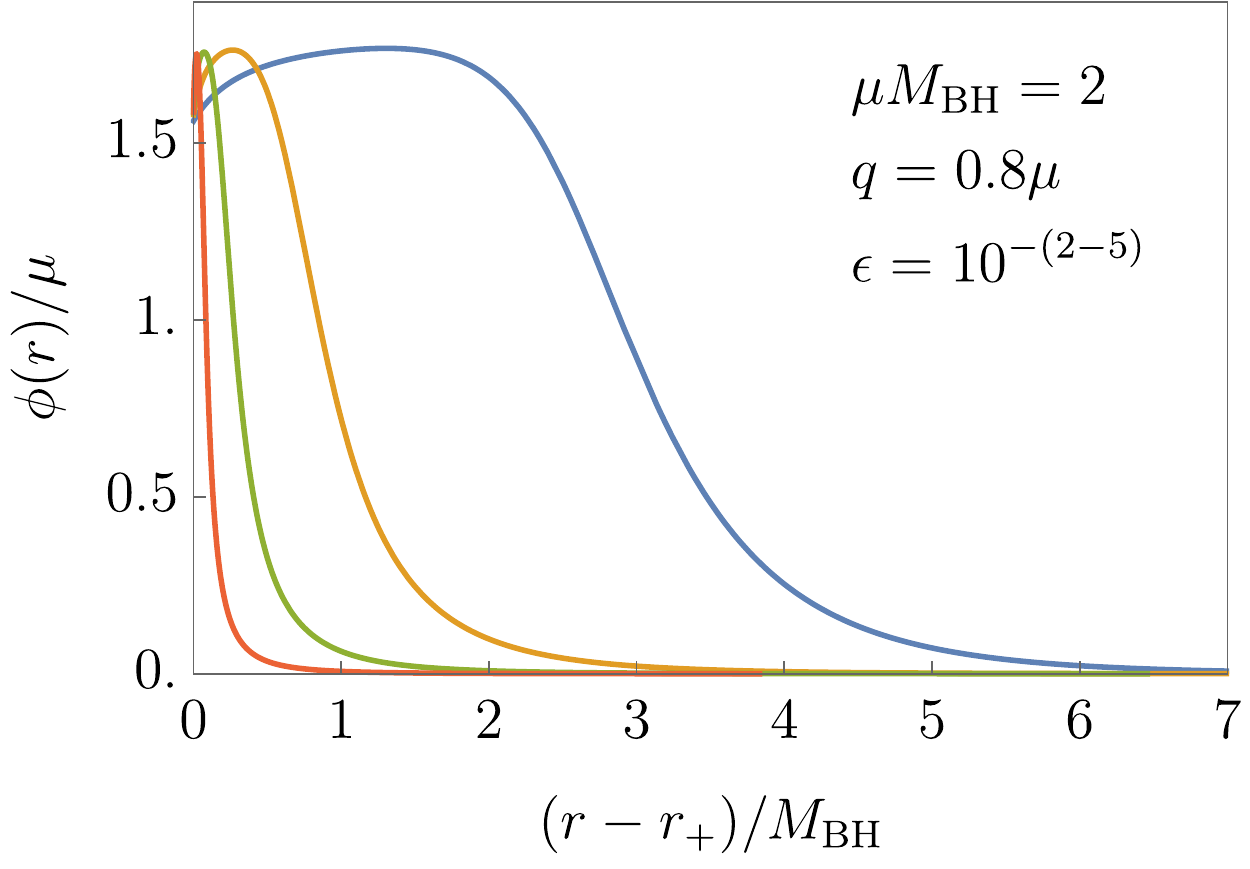}
 \caption{Examples of Q-ball solutions $\phi$ for near-extremal BHs. We take $\mu M_{\rm BH} = 2$, $q/\mu = 0.8$, 
 $Q_{\rm BH} = (1-\epsilon)M_{\rm BH}$ with 
 $\epsilon = 10^{-2}, 10^{-3}, 10^{-4}, 10^{-5}$ from right to left. }
\label{fig:prof}
\end{figure}

\begin{figure}
\centering
  \includegraphics[width=0.9\linewidth]{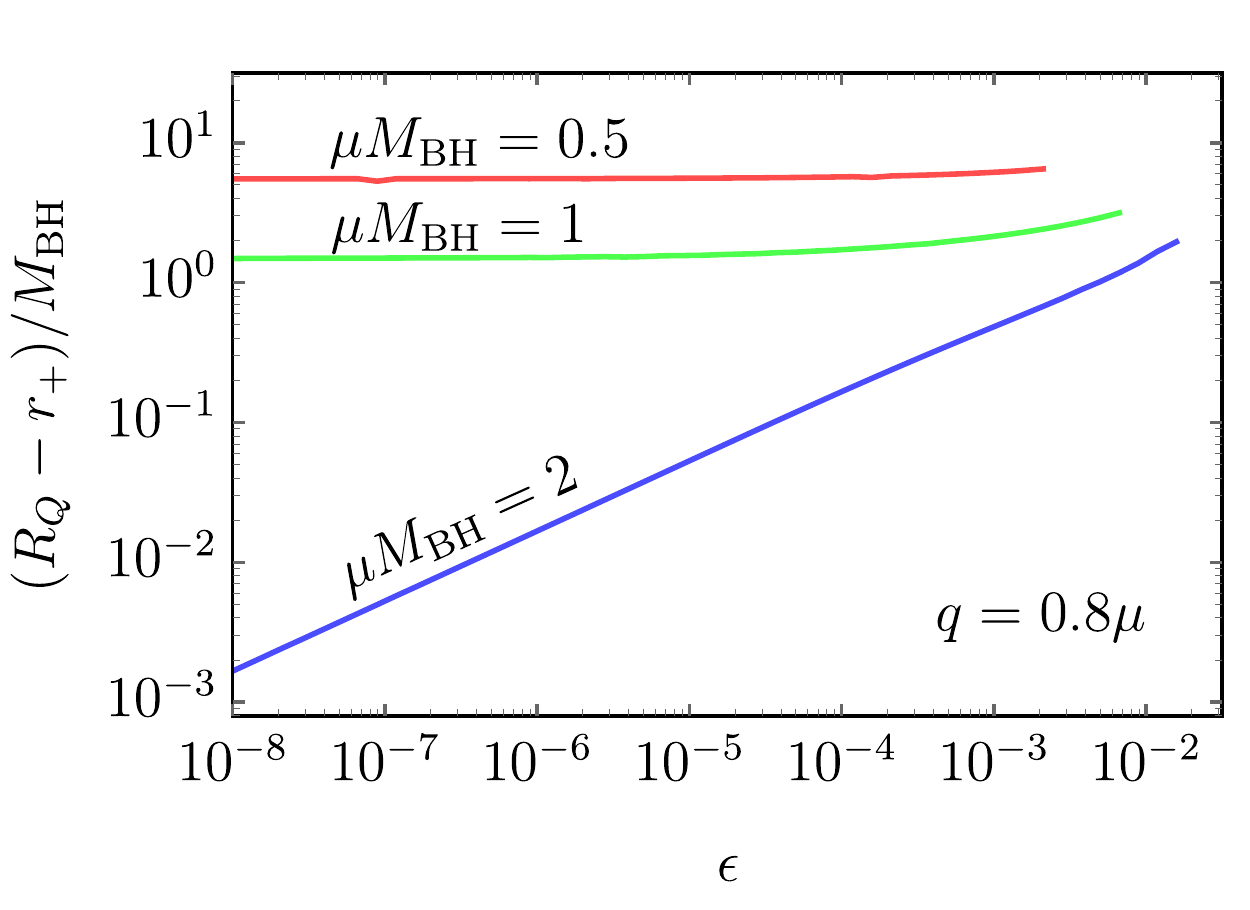}
 \caption{Q-ball width as a function of $\epsilon$. We take $q/\mu = 0.8$ and $\mu M_{\rm BH} = 0.5$ (red curve), $1$ (green curve), $2$ (blue curve). }
\label{fig2}
\end{figure}

From our numerical simulations, 
we find that Q-balls can only exist when $\mu \sqrt{1 - \frac{\lambda^2}{4 A \mu^2} } = \mu/2 \lesssim \omega_\phi \lesssim \mu$ for $M_{\rm BH} \mu \lesssim {\cal O}(1)$, which is consistent with the above discussion about large Q-clouds. Note that $\omega_\phi=\omega_c=qQ_{\rm BH}/r_+$ should hold for stability against the superradiance and $q/\mu$ and $Q_{\rm BH}/M_{\rm BH}$, which uniquely determine $\omega_c$, can be chosen to give $\omega_\phi$ in the above range. 
We also find that $\omega_\phi$ must be very close or equal to unity for $M_{\rm BH} \mu \gtrsim {\cal O}(1)$ and that no Q-cloud solution exists for $M_{\rm BH} \mu \gg 1$. 
For example, 
there is an upper bound by $\omega_\phi = 1$, at which $M_{\rm BH} \mu \simeq 8.0, 8.7, 12$ 
for the cases of $Q_{\rm BH}/M_{\rm BH} = 0.1, 0.5, 0.9$, respectively. 
Note that $q/\mu$ at this upper bound is determined by $\omega_c = \omega_\phi = 1$.

The possibility of the existence of Q-cloud around extremal BHs is an interesting question~\cite{Murata:2013daa, Angelopoulos:2018yvt}. 
However, the boundary conditions at the surface of an extremal BH do not uniquely determine a Q-cloud solution 
because the equation (\ref{eq}) is regular at $r=r_+$ only if $\phi (r_+) = \phi' (r_+) = 0$. 
Instead, we consider near-extremal BHs with $Q_{\rm BH} = (1-\epsilon) M_{\rm BH}$ for small values of $\epsilon$. 
Figure~\ref{fig:prof} shows Q-cloud profiles for the cases $\epsilon = 10^{-2}, 10^{-3}, 10^{-4}, 10^{-5}$, 
where we take $q /\mu = 0.8$ and $\mu M_{\rm BH}= 2$. 
We can see that the width of the Q-cloud, defined by $R_Q - r_+$, becomes small at the extremal limit of the BH. 
Fig.~\ref{fig2} plots the Q-cloud width as a function of $\epsilon$ for the cases of $q/\mu = 0.8$ and $\mu M_{\rm BH} = 0.5, 1, 2$. 
The width of the Q-cloud for the case of $\mu M_{\rm BH} = 2$ 
can be arbitrarily short in the extremal limit of BH. 
However, we note that this is not a generic feature for a relatively large $\mu M_{\rm BH}$. 
We find that the results are qualitatively different for the case of $q / \mu = 1$~\cite{futurework}.

The behavior of the width in the extremal limit of BH can be roughly understood by the following heuristic argument. 
Let us focus on the regime of $\sqrt{\epsilon} \ll (r - r_+)/r_+ \equiv x \ll 1$ and $\epsilon \ll 1$ 
so that we can approximate as $\Delta \simeq r_+^2 x^2$ 
and $r \simeq r_+$. 
Then the equation of motion is given by 
\beq
 \frac{\del^2 \phi}{\del x^2} + \frac{2}{x} \frac{\del \phi}{\del x} 
 + \frac{r_+^2 q^2}{x^2} \phi - \frac{r_+^2 V'}{x^2} = 0. 
 \label{conformal-eq}
\eeq
This equation respects the conformal symmetry under the approximation, namely it is invariant with respect to the rescaling of $x$. Therefore the typical size of the solution is not determined by this equation itself but is determined by the full equation beyond the ``ultraviolet" or ``infrared'' cutoffs of this equation. 
These cutoffs are given by ${\cal O}(\epsilon^{1/2})$ and ${\cal O}(1)$ because of the approximation we used to derive the equation. Thus we expect that the width of the Q-cloud
is proportional to either $\epsilon^{1/2}$ or $\epsilon^0$ for a small $\epsilon$, depending on the parameters. 
This is consistent with the results of our numerical calculation shown in Fig.~\ref{fig2}. 

In Appendix A, we consider the case in which the potential is given by a logarithmic function, motivated by the flat directions in gauge-mediated supersymmetric models~\cite{deGouvea:1997afu, Kusenko:1997si}. In this case, the Q-ball solution in flat spacetime is not a thin-wall type and scaling behaviors of parameters are different from the ones discussed above~\cite{Dvali:1997qv}. In particular, 
we find that a Q-cloud with an arbitrarily short thickness can be realized by taking a large value of $\omega_\phi$ (or $\mu$). 

\begin{figure}
\centering
  \includegraphics[width=0.9\linewidth]{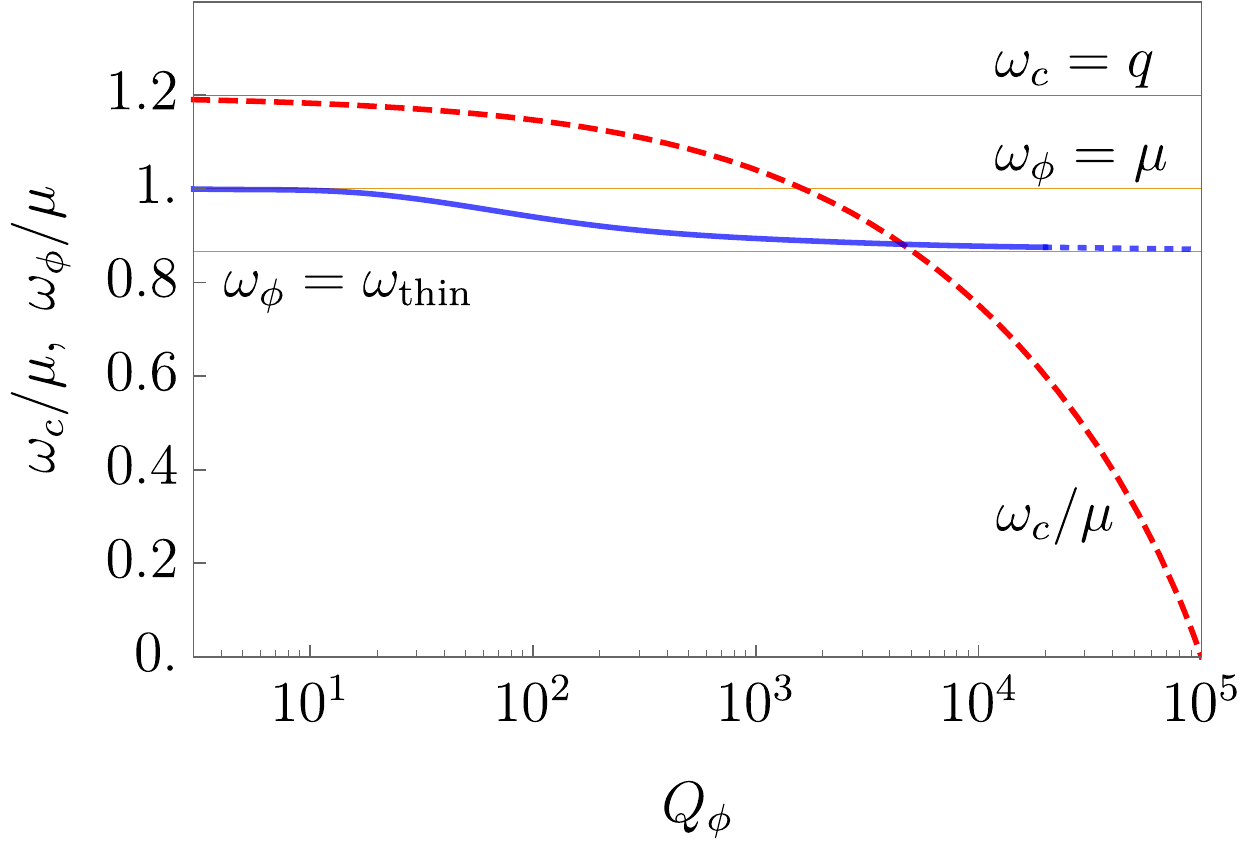}
 \caption{$\omega_i$-$Q_\phi$ relation for a fixed $Q_{\rm tot} = Q_\phi + Q_{\rm BH}$ ($= M_{\rm BH}$), where $i = \phi$ (blue solid curve) and $c$ (red dashed curve). We set $A/\mu = \lambda = 1$, $q /\mu = 1.2$ as an example. 
When $q > \mu$, which is suggested by the weak-gravity conjecture, the two lines $\omega_c$ 
and $\omega_\phi$ intersect with each other. 
To plot $\omega_\phi$, we use the polynomial potential with cubic and quadratic terms in flat spacetime as an example~\cite{Kusenko:1997zq}. 
This curve asymptotically approaches the value calculated by the thin-wall approximation, $\omega_{\rm thin}$, for $Q_\phi \to \infty$.}
\label{fig:omega}
\end{figure}

\vspace{0.1cm}
{\bf Evolution of Q-cloud.--}
Finally, we consider the evolution of a system with a fixed total charge $Q_{\rm tot} = Q_\phi + Q_{\rm BH}$. 
In the limit $R_Q \gg r_+$, we can approximate the Q-cloud solution from the solution in flat spacetime. 
Figure~\ref{fig:omega} shows a schematic of two curves: $\omega_\phi = \omega_\phi(Q_\phi) $ and $\omega_c = \omega_c (Q_{\rm BH})$ with fixed $Q_{\rm tot}$ ($= M_{\rm BH}$). 
If the blue solid curve ($\omega_\phi (Q_\phi)$ and the red dashed curve ($\omega_c (Q_{\rm BH})$) intersect, 
there exists a stationary solution of a BH with Q-cloud.
Note that stable or stationary Q-cloud do not exist by a continuous deformation from a BH without Q-cloud, 
so that in principle we cannot calculate $\omega_\phi$ for an arbitral value of $Q_\phi$. 
However, 
when $R_Q \gg R_{\rm BH}$ and $Q_\phi \ll Q_{\rm BH}$, 
the Q-cloud can be approximated by a Q-ball in flat spacetime 
and we can plot the curve $\omega_\phi = \omega_\phi(Q_\phi)$. 
This approximation is not justified for $Q_\phi \sim Q_{\rm BH}$, which 
is indicated by the blue dotted curve in the figure.

Note that under the superradiance condition, $i.e.$ $\omega_\phi<\omega_c$, 
the energy and charge of the BH will be extracted by the Q-cloud. 
This can also be understood by the fact that 
$\omega_c \equiv q Q / r_+$ is equal to the electric potential energy of particle with unit charge $q$ that comes from $r = \infty$ to the surface of the BH ($r= r_+$)~\cite{Christodoulou:1972kt} and hence we can identify $dE_{\rm BH} / d Q_{\rm BH} = \omega_c / q$. Since $d E_\phi / d Q_\phi = (1/q) d E_\phi / d N_\phi = \omega_\phi /q$, 
it is energetically favored for a charged particle to be extracted from the BH by the Q-cloud when $\omega_\phi < \omega_c$. 
As the Q-cloud extracts the charge of the BH, $\omega_c$ ($\propto Q_{\rm BH}$) decreases. 
On the other hand, if $\omega_\phi > \omega_c$, 
the charge of the Q-cloud is absorbed into the BH 
and $\omega_c$ increases. 
In both cases, $\omega_\phi$ changes only slightly, so 
the system will eventually reach the stationary solution of Q-cloud with $\omega_\phi = \omega_c$. This behavior demonstrates the stability of the Q-cloud if $\omega_\phi = \omega_c$. 
We also note that 
$\omega_c$ is always larger than $\omega_\phi$ for $Q_\phi$ being smaller than 
the critical point of $\omega_c = \omega_\phi$ for the case shown in Fig.~\ref{fig:omega}. 
This means that the Q-cloud is stable against even a large deformation. 
In particular, the Q-cloud with a BH inside is energetically favored compared to 
a BH that has consumed the entire Q-cloud (i.e., $Q_{\rm tot} = Q_{\rm BH}$). 
This ensures a stability of Q-cloud against quantum tunneling to a BH without Q-cloud. 
In Ref.~\cite{Sanchis-Gual:2015lje}, it is numerically demonstrated in a relativistic simulation 
that the same system but without the non-linear interaction reaches a stable hairy BH that exists at the threshold of the superradiant instability, if one sets a mirror~(box) outside the horizon. In our case, the non-linear self-interaction plays a similar role of the mirror because both prevents the scalar field from escaping. In this sense their results support our argument on the stability of the solutions we obtained. 

Next, we discuss the relation of this work to the weak-gravity conjecture. 
The conjecture states that there must exist a charged particle with mass $\mu$ and charge $q$ that satisfies 
$q > \mu$ in any gauge theories with gravity~\cite{ArkaniHamed:2006dz}. 
This must hold when the low-energy effective field theory comes from a consistent theory of quantum gravity, like string theory, since otherwise there does not exist a UV theory that contains quantum gravity. 
Note that $\omega_c$ is maximal and is equal to $q$ for an extremal BH (i.e., for $Q_{\rm BH} = M_{\rm BH}$). 
These facts imply that 
there must exist a particle that experiences superradiance 
around an extremal BH. 
Now consider a scalar field with self-interactions that allow it to form a Q-ball. 
This field would resemble the diagram in Fig.~\ref{fig:omega}. 
Then the weak gravity conjecture implies that 
$\omega_c \, (=q) > \omega_\phi \, (\simeq \mu)$ for $Q_\phi = 0$ and $Q_{\rm BH} = M_{\rm BH}$ 
while 
$\omega_c \, (=0) < \omega_\phi \, (\ne 0)$ for $Q_\phi = Q_{\rm tot} \, (=M_{\rm BH})$ and $Q_{\rm BH} = 0$. 
This statement means that 
the curves of $\omega_c$ and $\omega_\phi$ must intersect 
if we take $Q_{\rm tot} = M_{\rm BH}$. 
The existence of Q-cloud is thus supported by the weak-gravity conjecture 
if the scalar potential admits the formation of a Q-ball in flat spacetime.

\vspace{0.1cm}
{\bf Discussion.--}
We have shown that a Q-cloud solution exists that is narrower than the BH as the BH approaches the extremal limit. 
One may think that this is a counter example to the no short-hair theorem proven in Refs.~\cite{Nunez:1996xv, Hod:2011aa}. However, as noted in those papers, the charged BH does not satisfy one of the necessary condition of the theorem and the theorem is not applicable to our case. 
One may also think that 
our result is inconsistent with the conclusion of Ref.~\cite{Mayo:1996mv}, 
where they discuss that 
the RN BH cannot have a scalar hair even in the presence of a gauge field. 
In Appendix B, we argue that their argument cannot be applied if the scalar field as a nonzero mass term.
We also note that a counterexample has been found in Refs.~\cite{Kleihaus:2009kr, Kleihaus:2010ep}, which implies that 
a Q-cloud solution, which is a bound state of complex scalar field around a charged BH, 
is still allowed. 
In those papers they used a specific linear potential, which is singular at the origin, 
while we use a regular and more general potential to show the existence of the Q-cloud solutions.

Although charged BHs are unlikely to develop during the realistic evolution of cosmological history, the observational possibility of Q-cloud is still interesting to investigate. 
The recent observation of a BH shadow by the Event Horizon Telescope has introduced such a possibility~\cite{Akiyama:2019fyp}. 
The gravitational lensing of light may be affected by scalar cloud around a BH, and this effect could be observed by the Event Horizon Telescope~\cite{Falcke:1999pj, Cunha:2015yba}. 
If the scalar cloud has a certain type of photon coupling, we may observe photons that are produced from the accretion disc and are then polarized in interactions with the scalar cloud~\cite{Chen:2019fsq}. 
Although these works consider a Kerr BH, Q-cloud around a charged BH is also worth investigating.

Finally, we note that 
near-extremal Reissner-Nordstr\"{o}m BHs are well-studied in the context of string theory, 
as such an object can be described with a D-brane in certain spacetime dimensions~\cite{Strominger:1996sh, Callan:1996dv}. 
This kind of BH can be also studied using the anti-de Sitter/conformal field theory (AdS/CFT) correspondence, 
since the geometry around the near-extremal BH reduces to the AdS$_2 \times$S$^2$ geometry (see, e.g., Refs.~\cite{NavarroSalas:1999up, Guica:2008mu, Hartman:2008pb}). 
The properties of Q-cloud will also be interesting to explore using these approaches, and we leave this investigation for future work. 

\vspace{0.5cm}


{\bf Acknowledgement.--}
The work of M.S. is supported in part by a Research Fellowship for Young Scientists from the Japan Society for the Promotion of Science (JSPS).
M.S. appreciates the hospitality of the Tsung-Dao Lee Institute of the Shanghai Jiao Tong University where a part of this work was done.
J.P.H is supported by Korea NRF-2015R1A4A1042542. 

\vspace{0.5cm}

{\bf Appendix A: Case for a logarithmic potential.--}
In this Appendix, we consider the case in which the potential is given by a logarithmic function to study the thick-wall type solution. 
The potential is given by 
\begin{align}
 V(\Phi)= \mu^4 \ln \lmk 1 + |\Phi|^2 / \mu^2 \rmk. 
\end{align}
This is motivated by the flat directions in gauge-mediated supersymmetric models~[38, 43], where we rescale $\Phi$ to absorb a supersymmetric breaking scale without losing generality. 
The potential is almost flat for $\abs{\Phi} \gg \mu$.

In the limit of a large size of Q-cloud and small $q$, we can neglect the effects of both gravity and gauge potential 
and quote the results of Q-ball in the flat spacetime. 
In this case, $V(\phi) / \phi^2$ is minimized at an infinite value of $\phi$, which 
means that the solution is not a thin-wall type~[44]: 
\beq
 \omega_\phi \simeq \sqrt{2}\pi  Q_\phi^{-1/4} \mu \simeq \sqrt{2} \pi \mu^2 / \phi_0, 
 \qquad
 R_Q \simeq \frac{\pi}{\omega_\phi}, 
\eeq
where 
$\phi_0$ ($\propto Q_\phi^{1/4}$) is the field value at the center of the Q-ball 
and 
we neglect logarithmic corrections for the sake of simplicity. 
These scaling behavior are in contrast to the case of thin-wall type Q-ball like Eq.~(\ref{thin_wall_Q-ball}). 
In addition, $\omega_\phi$ can be made arbitrary small by taking a large value of $Q_\phi$. 
We can then construct a stationary solution of the Q-cloud for an arbitrary small $\omega_c$ by matching $\omega_\phi = \omega_c$. 
Note that the Q-ball radius is determined by $\omega_\phi$, which is the typical mass scale of the potential at $\phi = \phi_0$.

\begin{figure}
\centering
  \includegraphics[width=0.9\linewidth]{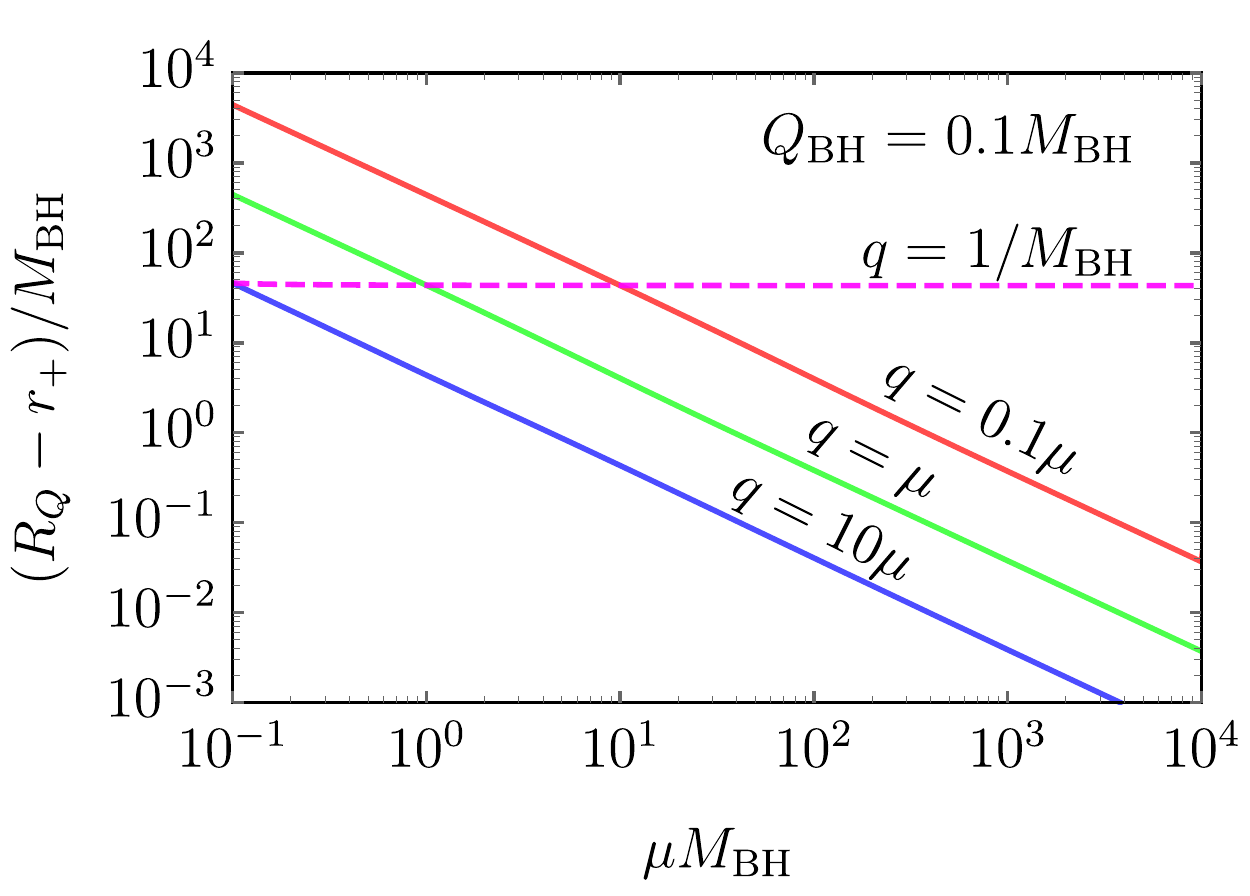}
 \caption{Q-ball width as a function of $M_{\rm BH} \mu$ for the logarithmic potential.
 We take $Q_{\rm BH} = 0.1 M_{\rm BH}$ and $q = 0.1 \mu$ (red curve), $\mu$ (green curve), $10 \mu$ (blue curve). 
 We take $q = 1/M_{\rm BH}$ for the dashed line. }
\label{fig:width}
\end{figure}

\begin{figure}
\centering
  \includegraphics[width=0.9\linewidth]{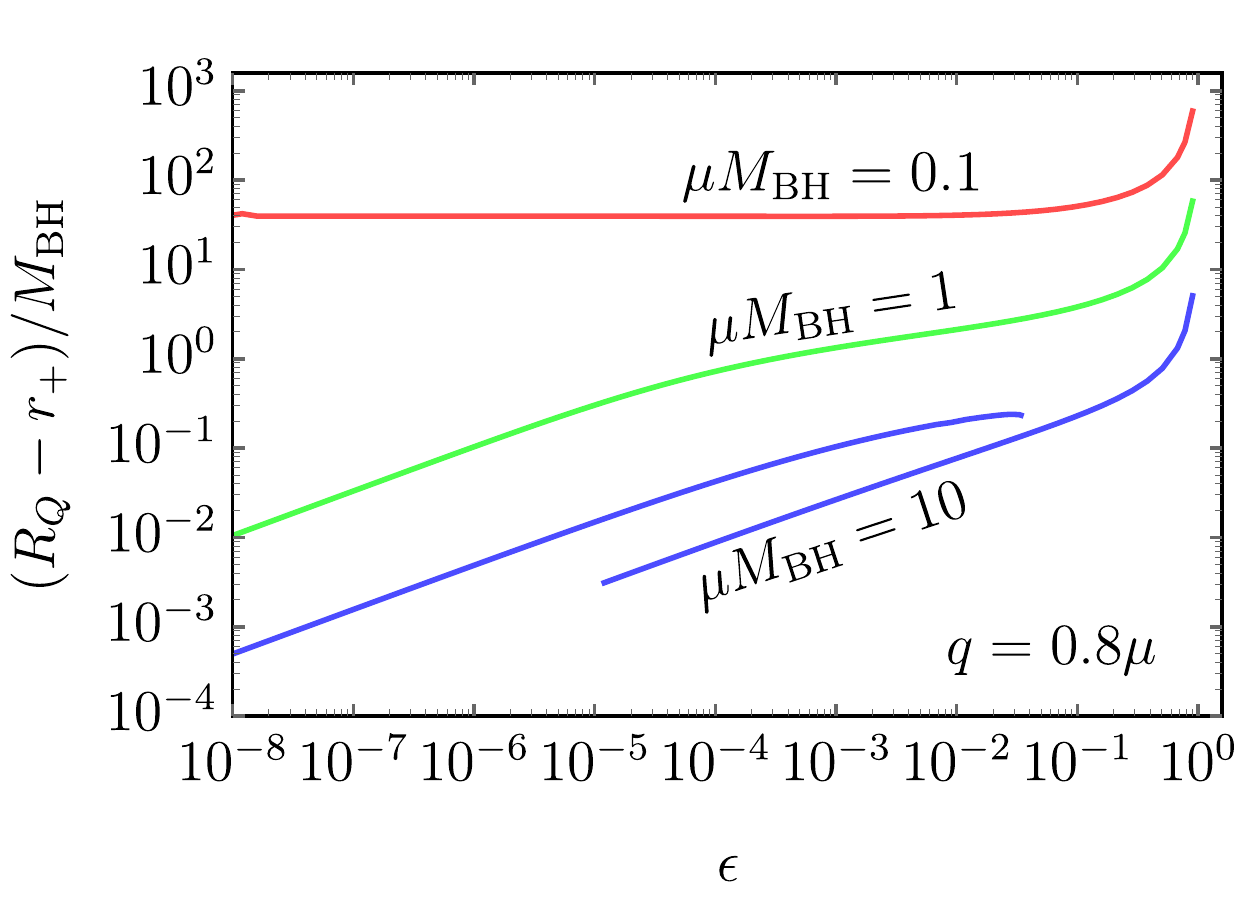}
 \caption{Q-ball width as a function of $\epsilon$ for the logarithmic potential. 
 We take $q/\mu = 0.8$ and $\mu M_{\rm BH} = 0.1$ (red curve), $1$ (green curve), $10$ (blue curve). 
 Two solutions of the Q-cloud exist for the case $\mu M_{\rm BH} =10$ and $10^{-5} \lesssim \epsilon \lesssim 4\times 10^{-2}$. 
 }
\label{fig4}
\end{figure}

When $R_Q \sim r_+$ and $Q_{\rm BH} \ll M_{\rm BH}$, 
$\omega_\phi$ ($= \omega_c$) remains as a parameter that determines the overall behavior of the solution to equation (7). 
Thus, we expect that the Q-cloud width, $R_Q - r_+$, is on the order of $1/ \omega_\phi$. 
Figure~\ref{fig:width} shows how the Q-ball width depends on $M_{\rm BH} \mu$ 
for the cases with $Q_{\rm BH} = 0.1 M_{\rm BH}$ and $q /\mu = 0.1, 1, 10$ (solid lines) and $q = 1/ M_{\rm BH}$ (dashed line). 
We find that $(R_Q - r_+) \sim \pi / \omega_\phi$ even for $R_Q - r_+ \ll r_+$ 
and that there exists a Q-ball solution for an arbitrarily large $M_{\rm BH} \mu$. 
Hence, we can construct a Q-cloud with an arbitrarily short thickness 
by taking a large value of $\omega_\phi$ (or $\mu$). 
This finding contrasts with the case of the polynomial potential, in which the Q-cloud cannot be much thinner than a non-extremal BH because of the upper bound on $M_{\rm BH} \mu$.

Figure~\ref{fig4} plots the width of the Q-cloud as a function of $\epsilon$ 
to show the Q-cloud behavior at the near-extremal limit. 
As in the case of the polynomial potential, 
the width can be arbitrarily short for a near-extremal BH in the limit of $\epsilon \to 0$ for the cases of $\mu M_{\rm BH} = 1$ and $10$. 
These results are consistent with our heuristic argument around Eq.~(\ref{conformal-eq}) 
because the width is proportional to either $\epsilon^{1/2}$ or $\epsilon^0$ for a small $\epsilon$. 
We also find that two solutions of the Q-cloud exist for the case of $\mu M_{\rm BH} = 10$ 
in the range of $10^{-5} \lesssim \epsilon \lesssim 4 \times 10^{-2}$.
We confirm that the solution on the upper branch has a smaller charge $Q_\phi$ compared to the one on the lower branch with the same $\epsilon$. This implies that the solution on the upper branch has a smaller $\omega_\phi$, hence a smaller energy, compared to the solution on the lower branch with the same $Q_\phi$. Therefore the solution on the upper branch is energetically more favored.

{\bf Appendix B: Consistency with no-hair theorem.--}

In this Appendix, we discuss that the no-hair theorem of Mayo and Bekenstein~\cite{Mayo:1996mv} cannot be applied to the case in which the complex scalar field has a non-zero mass and hence it does not contradict with the solutions we have found. 
The logic of the theorem that forbids the existence of the scalar-hair is as follows. 
From the equation of motion of the scalar field $\phi$ at $r\rightarrow\infty$ {\it without a scalar mass term}, one obtains an asymptotic solution as
\begin{align}
\phi \sim \frac1r\exp\left[i\sqrt{g(\infty)^2} \, r\right]. 
\end{align}
If $g(\infty)\neq0$, this leads to a divergent total charge since the charge density falls off too slowly at infinity. 
This therefore implies $g(\infty)=0$. 
Then, since $g(r)$ is monotonic, which can be proved from the equation of motion, 
$g(r)$ must be non-zero at the horizon. This then requires $\phi$ to be zero at the horizon, since otherwise the term $\sim (r^2/\Delta)^2 g^2 \phi$ in the equation of motion for $\phi$ will diverge. 
Finally, the authors show that the solution with $\phi(r_+)=0$ must be trivial (i.e., $\phi(r) = 0$ everywhere) since otherwise the energy-momentum tensor is infinite.

The crucial point of the above argument is that the authors implicitly omitted the mass term for $\phi$ at infinity. 
If the mass term is taken into account, the asymptotic solution becomes
\begin{align}
 \phi \sim \frac1r\exp\left[-\sqrt{m^2_\phi - g(\infty)^2} \, r\right],
\end{align}
where $m_\phi^2$ ($= V''(\phi)$) is the squared mass. 
This leads to a finite total charge even for nonzero $g (\infty)$ since the solution falls off exponentially. 
Then, the requirement $g(\infty)=0$, and hence $g(r_+)\neq0$ is relaxed and the solution with $g(\infty)\neq0$, $g(r_+)=0$ is allowed. 
Finally this allows $\phi$ with $\phi(r_+)\neq0$ because $g(r_+)=0$ safely makes $(r^2/\Delta)^2 g^2 \phi$ finite, provided that $g$ vanishes as fast as $\sim (r-r_+)$. We therefore conclude that a scalar-hair with $\phi(r_+)\neq0$ and $g(r_+)=0$ is allowed if the scalar has a non-zero mass term. 
We note that our numerical solutions actually satisfy the latter boundary conditions.

\bibliography{references}

\end{document}